\documentclass[12pt]{article}
\usepackage{graphicx}
\usepackage{amsmath} 
\usepackage{amssymb} 
\usepackage{amsfonts} 
\usepackage{rotating}
\usepackage{subfigure}
\usepackage{lscape}

\begin{document}

\title{Handling non-ignorable dropouts in longitudinal data: A conditional model based on a latent Markov heterogeneity structure}


\author{Antonello Maruotti\footnote{A. Maruotti -
              Southampton Statistical Sciences Research Institute, University of Southampton, 
              Emai:a.maruotti@soton.ac.u.uk}           
}

\date{}

\maketitle

\begin{abstract}
We illustrate a class of conditional models for the analysis of longitudinal data suffering attrition in random effects models framework, where the subject-specific random effects are assumed to be discrete and to follow a time-dependent latent process. The latent process accounts for unobserved heterogeneity and correlation between individuals in a dynamic fashion, and for dependence between the observed process and the missing data mechanism.  Of particular interest is the case where the missing mechanism is non-ignorable. To deal with the topic we introduce a conditional to dropout model. A shape change in the random effects distribution is considered by directly modeling the effect of the missing data process on the evolution of the latent structure. To estimate the resulting model, we rely on the conditional maximum likelihood approach and for this aim we outline an EM algorithm. The proposal is illustrated via simulations and then applied on a dataset concerning skin cancers. Comparisons with other well-established methods are provided as well.
\end{abstract}

\section{Introduction}
\label{intro}

In longitudinal studies, measurements on the same subjects are taken repeatedly over time. Longitudinal data models permit tracing the dynamics of behaviors. Factors influencing these dynamics and unobserved characteristics driving intra-subjects correlations can be identified. 

Nevertheless, longitudinal data may suffer from missingness. This means that subjects may not be measured in some of the planned occasions, or exit the study at a given time point before its completion. The central concern in the analysis of longitudinal data suffering missingness is selection bias, that is a distortion of the estimation results due to non-random patterns of missingness.  Literature on missing data has so far been based on specific definition of the missing process. According to Rubin's taxonomy (Rubin, 1976), we define missing data as missing completely at random (MCAR) when the distribution of the missing data process do not depend either on the observed or unobserved outcomes, missing at random (MAR) when the missingness depends on the observed information, missing not at random (MNAR) when the missing data process is assumed to be related with both the observed and the unobserved responses. A further taxonomy could be specified and of interest according to the influence of missingness inspected with respect to parameter estimates. A missing data process is ignorable when a combination of MAR and separability in model parameters between the response and the missing process hold, and is non-ignorable (or informative) when a MNAR mechanism hold. Interesting reviews on missing data models are provided by Fitzmaurice (2003); Jansen et al. (2006); Diggle et al. (2007); Ibrahim and Molenberghs (2009).

A common approach to deal with non-ignorable dropouts is via a shared parameter approach (Wu and Carroll, 1988), where the observed process and the missingness mechanism share the same, continuous (Follmann and Wu, 1995) or discrete (Alf\`{o} and Aitkin, 2000; Tsonaka et al., 2009; Jung et al., 2011; Maruotti, 2011a; Belloc et al., 2012), latent structure. A limitation of shared parameter models is that the latent structure, often modeled via individual-specific random effects, is time-constant. Thus, the effect of unobserved heterogeneity is constrained to be constant over time or to evolve over time along a pre-specified parametric form (e.g., linear through inclusion of a random slope). 

In order to overcome these limitations, we develop a new approach to explore how inferences may change if the assumption of MAR is violated, by introducing a latent structure based on a latent Markov chain. We propose a shared parameter model, in which the unobserved heterogeneity is assumed discrete and modeled through the inclusion of random effects, whose distribution follow a first-order Markov chain (see e.g. Bartolucci et al., 2012; Maruotti 2011b; Bartolucci and Farcomeni, 2009). A conditional to drop-out model is introduced and proper inference is conducted. A shape change in the random effects distribution is considered by directly modeling the effect of the missing data process on the evolution of the latent structure. The observed measurements are modeled through a generalized linear mixed model. This brings several appealing features. Apart from providing a more flexible modeling tool, the dependency of missingness on the latent structure is assumed to vary over time. A strong advantage, over existing shared parameters models, is that we are able to cluster subjects into latent groups at different times and estimate the evolution of class changes over time. Furthermore, although computational burden increases with respect to standard shared parameters models, fitting a random effects model based on a hidden Markov structure is stable and acceptable in terms of computational time. Of course, this is not the only attempt to deal with time dependence under a longitudinal setting with non-ignorable missingness. Our proposal is somehow related to the model proposed by Albert and Follman (2007), where a continuous time-varying latent variable is considered, at the cost of a cumbersome computational effort. Assuming a discrete instead of a continuous latent process also has the advantage of permitting exact computation of the likelihood of the model without requiring quadrature or Monte Carlo methods. Another related approach is provided by Spagnoli et al. (2011) in a hidden Markov model framework. It is potentially very useful, but, as the authors stated, may not always be appropriate due to specific constraints imposed.

For the maximum likelihood estimation of the proposed model, we use an EM-based algorithm deriving, and slightly modifying, recursions from the Baum-Welch algorithm (Baum et al. 1970; Welch, 2003), widely adopted in the hidden Markov models literature (see e.g. Zucchini and MacDonald, 2009; Chapter 4; Bartolucci et al., 2012; Chapter 3). However, different methods can be used in order to provide parameter estimates (see e.g. Bulla and Berzel, 2008).

We illustrate the proposal by a simulation study in order to investigate the empirical behavior of the proposed approach with respect to several factors, such as the number of observed units and times, providing a comparison with well-established approaches. Finally, the proposed approach is illustrated by a dataset of skin cancer results from patients in a clinical trial concerning the effects of $\beta$-carotene to prevent non-melanoma skin cancer Greenberg et al. (1990). Previous analysis on these data focused on the modeling of dropout process and showed substantial heterogeneity among patients as well as the importance of accounting for the missingness mechanism, related with the time in the treatment (Hasan et al., 2009; Maruotti, 2011a).

The article is organized as follows. In Section 2, we briefly review relevant aspects necessary for the introduction of our approach. In Section 3, we illustrate the proposed model for non-ignorable missingness. Likelihood inference is provided in Section 4. In Section 5 we show the results of a simulation study by analyzing the behavior of the maximum likelihood estimator under different simulation schemes. 
The application to the skin cancer dataset is illustrated in Section 6, along with model selection. Some remarks along with drawbacks, which may arise by adopting the proposed methodology, and future research are discussed in Section 7.

\section{Preliminaries}
\label{prel}

Let $Y_{it}$ be a response variable recorded on $n$ subjects ($i=1,2,\dots,n$) at $T$ scheduled times ($t = 1,2,\dots,T$), together with a set of $p$ covariates $ {\bf x}_{it}=\{x_{it1},x_{it2},\dots,x_{itp}\}'$. Let us decompose
the design vector as $ {\bf x}_{it} = \{{\bf x}_{1it},{\bf x}_{2it}\}$, where variables whose effects are assumed to be fixed across
subjects are collected in ${\bf x}_{1it}$ , while those which vary over classes are in ${\bf x}_{2it}$,  
\begin{equation}
g(E[Y_{it}\mid {\bf u}_i,{\bf x}_{1it},{\bf x}_{2it}]) = {\bf x}_{1it}'\boldsymbol{\beta}+{\bf x}_{2it}' {\bf u}_i
\end{equation}

\noindent where $g(\cdot)$ is a strictly monotone link function, $\boldsymbol{\beta}$ is a vector of fixed parameters and ${\bf u}_i$ is a vector of individual-specific random effects accounting for unobserved heterogeneity between individuals in the regression parameters and assumed i.i.d. with a common but unknown density function. 

Nevertheless, some subjects may leave the study before its completion time, thus presenting incomplete data records. Let ${\bf R}_i = \{R_{i1},R_{i2},\dots,R_{iT}\}$ be a vector of missingness indicators. Each element of ${\bf R}_i$ takes value zero if the corresponding value of ${\bf Y}_i = \{Y_{i1},Y_{i2},\dots,Y_{iT}\}$ is observed and the value of one if the corresponding value of ${\bf Y}_i$ is missing. For the special case of dropout, the missing data pattern is monotone and can be summarized by $S_i = T-\sum_{t=1}^T R_{it}$. 
We assume that form some link function $g(\cdot)$: 
\begin{equation}
g(E[S_{i}\mid {\bf u}_i,{\bf v}_{1i},{\bf v}_{2i}]) = {\bf v}_{1i}'\boldsymbol{\tilde{\beta}}+{\bf v}_{2i}' {\bf u}_i
\end{equation}
\noindent where $\boldsymbol{\tilde{\beta}}$ is a vector of parameter and $ {\bf v}_{i} = \{{\bf v}_{1i},{\bf v}_{2i}\}$ is a vector of covariates.

Note that $S_i$ is assumed independent of $Y_{it}$ conditionally on ${\bf u}_i$.
Given these assumptions, the values of $Y_{it}$ and $S_i$ are linked by ${\bf u}_i$ and their joint distribution represents the so-called shared parameter model defined as

$$f(y_{i1},\dots,y_{iS_i},s_i)=\int\prod_{t=1}^{S_{i}}f(y_{it}\mid {\bf x}_{it}, {\bf u}_i)f(s_i\mid {\bf v}_{i}, {\bf u}_i)f({\bf u}_i)d{\bf u}_i.$$

However, inferences can be highly sensitive to misspecification of $f(s_i\mid {\bf v}_{i}, {\bf u}_i)$. An alternative to the specification of the dropout model has been introduced by Follmann and Wu (1995) and focuses on the conditional distribution $f(y_{i1},\dots,y_{iS_i}\mid s_i)$. Thus, the previous expression could be re-expressed as 

$$f(y_{i1},\dots,y_{iS_i},s_i)=\int\prod_{t=1}^{S_{i}}f(y_{it}\mid {\bf x}_{it}, {\bf u}_i)f(s_i\mid {\bf v}_{i})f({\bf u}_i\mid s_i)d{\bf u}_i$$

or, equivalently

$$f(y_{i1},\dots,y_{iS_i}\mid s_i)=\int\prod_{t=1}^{S_{i}}f(y_{it}\mid {\bf x}_{it}, {\bf u}_i)f({\bf u}_i\mid s_i)d{\bf u}_i.$$

\noindent where the conditional distribution $f({\bf u}_i\mid s_i)$ is different from the marginal random effects distribution $f({\bf u}_i)$.

The resulting conditional log-likelihood has the following form
$$\ell(\cdot\mid S_i) = \sum_{i=1}^n\log\left\{\int\prod_{t=1}^{S_{i}}f(y_{it}\mid {\bf x}_{it}, {\bf u}_i)f({\bf u}_i\mid s_i)d{\bf u}_i\right\}.$$

As pointed out by Alf\'o and Aitkin (2000) and Tsonaka et al. (2009), important drawbacks may arise if a parametric (e.g. Gaussian) distribution for the random terms is assumed. A theoretically more appealing alternative is the nonparametric
maximum likelihood (NPML) estimation of the mixing distribution. Thus, relying on the results of Laird (1978) and Lindsay (1983a, 1983b), the NPML of the mixing distribution is a discrete distribution on a finite number of mass points, say $J$.
The conditional log-likelihood becomes equal to
$$\ell(\cdot\mid S_i) = \sum_{i=1}^n\log\left\{\sum_{j=1}^J\prod_{t=1}^{S_{i}}f(y_{it}\mid {\bf x}_{it}, {\bf u}_i={\bf u}_j)f({\bf u}_i={\bf u}_j\mid s_i)\right\}$$

where $$f({\bf u}_i={\bf u}_j\mid s_i) = \pi_{ij} = \frac{\exp(\gamma_{0j}+\gamma_{1j}s_i)}{1+\sum_{j=1}^{J-1}\exp(\gamma_{0j}+\gamma_{1j}s_i)}$$
\noindent and ${\bf \gamma}_j = \{\gamma_{0j},\gamma_{1j}\}$ is a vector of fixed regression parameters. Of course ${\bf \gamma}_{J}=\{\gamma_{0J},\gamma_{1J}\}=0$  to ensure identifiability.

The main idea behind this approach is that a latent variable exists, stratifies the data in a certain number of groups and accounts for dependence between the response variable and the missing data process. The missing data process is assumed to be related with the unknown class membership via a regression model.

\section{A time-varying random effects model for incomplete data}
\label{models}
In discrete random effects models, class memberships, used to take into account the unobserved heterogeneity between subjects, are assumed time-constant. This assumption is common to many models for longitudinal data, even to the shared parameter models. However, time-varying memberships could be considered (see e.g. Bartolucci and Farcomeni, 2009). Indeed, if the effect of unobservable factors on
the responses of a subject is not time-constant, there can be bias in the parameter estimates.

The temporal evolution of state membership can be conveniently described by including a vector of time-varying
random effects, say ${\bf u}_{it}$. Thus, regarding the distribution of the subject-specific parameters, for each subject $i$ the random vector ${\bf u}_{it}$ is assumed to follow a first-order Markov chain with states ${\bf u} = \{{\bf u}_1,{\bf u}_2,\dots,{\bf u}_J\}$ with initial probabilities $\pi_{ij}(S_i)= \Pr({\bf u}_{i1}={\bf u}_{j}\mid S_i)$, $j=1,2,\dots,J$, $\sum_{j=1}^J\pi_{ij}(S_i)=1$, and transition probabilities $\pi_{ikj}(S_i) = \Pr({\bf u}_{it}={\bf u}_{j}\mid {\bf u}_{it-1}={\bf u}_{k}, S_i)$, $j,k = 1,2,\dots,J$, $\sum_{j=1}^J\pi_{ikj}(S_i)=1$. 

To take the missingness issue into account the distribution of the latent structure defined is allowed to depend on the time to dropout. Note that the initial and transitions probabilities of the latent process are independent of any other covariates. This assumption could be easily relaxed (see e.g. Maruotti and Rocci, 2012). The presence of conditioning, causing a change in the distribution of the latent process, may be treated through the specification of a conditional model for the latent process and $S_i$. 

In the spirit of Alf\`{o} and Aitkin (2000), the probability of being in a given state at time 1 can be determined though the following regression model

\begin{equation}\label{delta0}
\pi_{ij}(S_i)= \frac{\exp(\gamma_{0j}+\gamma_{1j}s_i)}{1+\sum_{h=1}^{J-1}\exp(\gamma_{0h}+\gamma_{1h}s_i)}
\end{equation} 

\noindent and similarly we link the time to dropout and the entries of the transition probability matrix capturing the evolution over time as 

\begin{equation}\label{trans0}
\pi_{ikj}(S_i)= \frac{\exp(\phi_{0kj}+\phi_{1kj}s_i)}{1+\sum_{h=1}^{J-1}\exp(\phi_{0kh}+\phi_{1kh}s_i)}.
\end{equation}

Subjects do not share the same latent structure, except if the have the same value of $S_i$. Indeed, different homogeneous (over time) Markov chain have been defined conditionally on $S_i$. 

\noindent The specification of the model is completed by assuming that 
\begin{equation}\label{observed}
g(E[Y_{it}\mid {\bf u}_{it},{\bf x}_{1it},{\bf x}_{2it}]) = {\bf x}_{1it}'\boldsymbol{\beta}+{\bf x}_{2it}' {\bf u}_{it}
\end{equation}

The model specified by (\ref{delta0}), (\ref{trans0}) and (\ref{observed}) introduces more flexibility with respect to commonly used shared parameter models, relaxing the assumption that the observations are conditionally independent over time given the latent class. A key assumption (unverifiable, but common) implied by our proposal is that within latent states, the probability of observing  dropouts does not depend on missing data, after conditioning on the observed data. We would remark that, in some cases, the probabilistic structure defined so far can be used only for the purpose of computing some statistical estimates and inferences which may or may not reflect the reality. On the other hand, the inclusion of a time-dependent structure is sensible in many applied context, especially in the presence of {\it long} panels and/or with a limited and/or time-constant set of observable covariates (see e.g. Wall and Li, 2009; Maruotti and Rocci, 2012; Delattre and Lavielle, 2012).

\section{Likelihood inference}
\label{likeinf}
Inference for the proposed model is based on the log-likelihood

$$\ell(\boldsymbol{\theta}) = \sum_{i=1}^n\log [f(y_{i1},\dots,y_{it}\mid S_i)]$$

where $\boldsymbol{\theta}$ is short-hand notation for all the nonredundant model parameters.
According to model assumptions $f(y_{i1},\dots,y_{it}\mid S_i)$ is equal to

$$\sum_{{\bf u}_{i1}}\cdots \sum_{{\bf u}_{iT}}\left[\Pr({\bf u}_{i1}\mid S_i)\prod_{t=2}^T\Pr({\bf u}_{it}\mid {\bf u}_{it-1}, S_i)\prod_{t=1}^Tf(y_{it}\mid {\bf u}_{it})\right]$$
with the sum $\sum_{{\bf u}_{it}}$
extended to all the possible configurations of ${\bf u}_{it}$.

\subsection{Estimation}
In order to estimate ${\boldsymbol \theta}$, we maximize $\ell(\boldsymbol{\theta})$ by using an EM-based algorithm. The algorithm is based on the definition of the so-called complete-data log-likelihood function, obtained by considering the sampling distribution of both the observed and the unobserved quantities. As our model is a mixture, the unobserved quantities are not only the missing measurements, but also the unknown states memberships. Treating these quantities as missing values, reflecting different sources of incomplete information, we define the complete-data log-likelihood function as

\begin{eqnarray}\label{complete}
\ell_{comp} &=& \sum_{i=1}^n\sum_{j=1}^J \xi_{i1j}\log\pi_{ij}(S_i) \nonumber\\
&+& \sum_{i=1}^n\sum_{t=2}^{S_i}\sum_{j=1}^J\sum_{k=1}^J\zeta_{itkj}\log\pi_{ikj}(S_i)\nonumber\\
&+& \sum_{i=1}^n\sum_{t=1}^{S_i}\sum_{j=1}^J\xi_{itj}\log f(y_{it}\mid{\bf x}_{1it},{\bf x}_{2it},{\bf u}_{it}).
\end{eqnarray}

where the variable $\zeta_{itkj} = I({\bf u}_{it}={\bf u}_{j},{\bf u}_{it-1}={\bf u}_{k})$ is an indicator variable equal to 1 if unit $i$ belongs to state $j$ at time $t-1$ and to state $k$ at time $t$, and $\xi_{itj}=I({\bf u}_{it}={\bf u}_{j})$ equals 1 if unit i at time $t$ belongs to state $j$ and 0 otherwise.

The algorithm is iterated by alternating the expectation (E) and the maximization (M) steps. Given the estimate $\hat{\boldsymbol{\theta}}_r$ obtained at the $r$-th iteration, the $(r+1)$-th iteration is initialized by an E-step, which evaluates the expected values of (\ref{complete}) with respect to the conditional distribution of the missing values given the observed data

\begin{eqnarray}
Q(\boldsymbol{\theta}\mid\hat{\boldsymbol{\theta}}_r) &=& \sum_{i=1}^n\sum_{j=1}^J\hat{\xi}_{i1j} \log\pi_{ij}(S_i) \nonumber\\
&+& \sum_{i=1}^n\sum_{t=2}^{S_i}\sum_{j=1}^J\sum_{k=1}^J\hat{\zeta}_{itkj}\log\pi_{ikj}(S_i)\nonumber\\
&+& \sum_{i=1}^n\sum_{t=1}^{S_i}\sum_{j=1}^J\hat{\xi}\log f(y_{it}\mid {\bf x}_{1it},{\bf x}_{2it},{\bf u}_{it}).
\end{eqnarray}

where $\hat{\xi}_{itj} = {\mathbb E}(\xi_{itj}\mid {\bf y}_i; \hat{\boldsymbol{\theta}}_r)$ and $\hat{\zeta}_{itkj} = {\mathbb E}(\zeta_{itkj}\mid {\bf y}_i; \hat{\boldsymbol{\theta}}_r)$.

To compute the expected complete log-likelihood, given the observed data and a set of parameter estimates, for each unit $i$, we should compute the posterior probabilities for all $J^T$ possible trajectories
through the hidden space. We address this computational issue through the Baum-Welch algorithm (for
a brief review, see e.g. Maruotti, 2011b). It provides the relevant marginal posterior probabilities $\hat{\xi}_{itj}$ and $\hat{\zeta}_{itkj}$ without
calculating the posterior probabilities of all possible trajectories through the hidden space. This is carried out using a set of recursive formulas, yielding a method for which the computational complexity
of the problem increases only linearly with the number of time points. 
In our approach, we partition the M-step in three subproblems, where the expected complete log-likelihood is maximized with respect to a subset of parameters given the current values of the others.
This leads to a (local) maximum because each subproblem is mathematically independent from the
other. In particular, the maximum with respect to $\pi_{ij}$ and $\pi_{ikj}$ is obtained as solutions of the following M-step equations, respectively,

\begin{equation}\label{delta}
\sum_{i=1}^n\sum_{j=1}^J\hat{\xi}_{i1j}\frac{\partial\log\pi_{ij}}{\partial\boldsymbol{\gamma}_j}=0
\end{equation}

and

\begin{equation}\label{trans}
\sum_{i=1}^n\sum_{t=2}^T\sum_{k=1}^J\hat{\zeta}_{itkj}\frac{\partial\log\pi_{ikj}}{\partial\boldsymbol{\phi}_{kj}}=0
\end{equation}

which are weighted sums of $J$ equations with weights $\xi_{i1j}$ and $\zeta_{itkj}$, respectively. Similarly,the updated estimates of $\boldsymbol{\beta}$ and ${\bf u}_{j}$ are computed simultaneously. Defining ${\bf \lambda}^* = \{\beta_1,\dots,\beta_p,{\bf u}_1,\dots,{\bf u}_J\}$, we obtain the updated estimates solving the following equation

\begin{equation}
\sum_{i=1}^n\sum_{t=1}^T\sum_{j=1}^J\hat{\xi}_{itj}\frac{\partial\log f(y_{it}\mid {\bf x}_{1it},{\bf x}_{2it},{\bf u}_{it}={\bf u}_{j})}{\partial\boldsymbol{\lambda}^*}=0
\end{equation}

To avoid the multinomial regressions defined by (\ref{delta}) and (\ref{trans}), we could exploit the $S_i$'s discrete nature as described in the Appendix. 
\subsection{Computational aspects}
The EM algorithm may converge to local maxima of the log-likelihood function. The presence of multiple local maxima is well documented in the case of general latent class models. A number of initialization strategies could be pursued (see for a general discussion e.g. Scharl et al., 2010). To avoid local maxima, we follow short runs of EM. The algorithm  is run several times from random starting points before passing to the EM algorithm without waiting for convergence using the threshold value 

\begin{equation}
\frac{|\log L_{r+1}-\log L_r|}{|\log L_r|}<10^{-2}.
\end{equation}

We have observed that convergence to spurious maxima could be detected using short EM runs by monitoring class proportions. We selected the ten outputs of the EM short run maximizing the log-likelihood and checked
for spurious solutions, where this effect did not occur. Then, these ten parameter sets were
used to initialize longer runs of the EM algorithm. The final convergence criterion for the EM is set to 

\begin{equation}
\frac{|\log L_{r+1}-\log L_r|}{|\log L_r|}<tol,
\end{equation}
where the tolerance $tol$ is set to $10^{-5}$.

Two drawbacks arise in the estimation procedure. The EM algorithm typically requires on inconveniently large number of iterations. To overtake this issue, the final steps of the algorithm could be replaced by a direct maximization procedure (using a Newton-type or Nelder-Mead simplex algorithms), which seems to work well and to be numerically stable when initial parameters are in the neighborhood of the maximum (see Bulla and Berzel, 2008; Zucchini and MacDonald, 2009, Chap. 3). Direct numerical maximization has appealing properties, especially concerning flexibility in fitting complex models and
the speed of convergence in the neighborhood of a maximum. The main disadvantage
of this method is its relatively small circle of convergence. Up to our experience, we suggest to start the estimation procedure with
the EM algorithm and switches to a Newton-type algorithm when a certain stopping
criterion is fulfilled. This leads to an algorithm that yields the stability and large circle of convergence from the EM algorithm along with
superlinear convergence of the Newton-type algorithm in the neighborhood of the
maximum. Since direct numerical maximization procedures implemented in general software can often perform
unconstrained numerical minimization only, the parameter constraints need
to be taken into account by different transformation procedures.

 Furthermore, the EM algorithm outlined above does not produce standard errors of the estimates, because approximations based on observed information matrix often requires a very large sample size. Thus, to obtain standard errors, we consider a parametric bootstrap approach, refitting 200 bootstrap samples simulated from the estimated model parameters. The approximate standard error of each model parameter is then computed.

\section{Simulation study}
\label{simula}
We carried out a simulation study to investigate the performance of the proposed approach focusing on the potential loss in efficiency resulting from latent structure misspecification. We generated several samples under different data generating mechanisms with the observed process given by a binary variable. We planned the simulation study to cover longitudinal schemes with different null dropout models. Data are simulated according to latent Markov heterogeneity structure. For each null model, we simulated 200 samples considering two experimental factors: the number of analyzed subjects $n= 100; 250; 500$ and the number of  maximum repeated measurements $T=5;10$; we fix the number of latent states $J=2$. We estimated the parameter values of two different models on each sample: a conditional model based on a latent Markov heterogeneity structure, where the dropout mechanism modifies the shape of the latent distribution (which in the following will be referred to as $M1$); and a conditional model with discrete random effect ($M2$), where the latent structure is assumed time-constant, i.e.ignores the presence of time-dependence (as described in Alf\'o and Aitkin, 2000).

\subsection{Conditional simulation scheme}

The dropout indicator variable $S_i$ was drawn from a discrete distribution, with $\Pr(S_i=t)=1/T, t = 1,\dots,T$ . Thus, on average (T-1)/T\% of subjects dropout prematurely. The latent process is defined according to (\ref{delta})-(\ref{trans}). The following set of parameters
$$
\boldsymbol{\gamma} = 
\left[\begin{array}{c}
\gamma_{01}\\
\gamma_{11}
\end{array}\right] = 
\left[\begin{array}{cc}
2.0\\ -0.5
\end{array}\right]\quad
\boldsymbol{\phi}_{11} = 
\left[\begin{array}{c}
\phi_{011}\\
\phi_{111}
\end{array}\right] = 
\left[\begin{array}{cc}
5.0\\ -1.5
\end{array}\right]\quad
\boldsymbol{\phi}_{21} = 
\left[\begin{array}{c}
\phi_{021}\\
\phi_{121}
\end{array}\right] = 
\left[\begin{array}{cc}
5.0\\ -0.75
\end{array}\right]\quad
$$
was chosen so that latent process was strongly related to dropout time. 

We consider a single covariate linear predictor where $x_{it}$ was drawn from a $N(0,1)$ distribution with fixed parameter $ \beta = 0.5$ and state-specific intercepts equal to $1.0$ and $-1.5$ respectively.

To evaluate model performance in recovering the true model structure and to investigate the effects of misspecification in drawing inference, we compared models on the basis of their ability to accurately estimate the fixed effect. We measured the accuracy by evaluating bias, standard errors and mean square errors of the estimators. We started each run from a randomly chosen set of parameters and stopped the algorithms when the increase in the log-likelihood was less than $10^{-5}$. For each of the two models, whenever possible, we used the same starting points. The experiments provide evidence to draw some conclusions. We summarize simulations results in Table 1.

The conditional model based on a latent Markov heterogeneity structure perform well, in terms of both bias and standard errors. The gain obtained by such a model with respect to its correspondent (more parsimonious) time-constant specification is sensible; the use of a more complex model provides significant improvements (see Table 1). This show that a misspecification of the latent structure may cause a severe bias in the estimation of the observed process, even if the latter is correctly specified, confirming the results of Maruotti and Rocci (2012). Moreover, when simulation study design allows for $T=5$ and $n=100; 250$ repeated measurements, the more parsimonious time-constant approach provides the same performance (even slightly better) than the {\it true} model. By increasing the length of subjects' sequence leads to an improvement in the parameter estimates, as expected. 

In conclusion, the simulation study provide insights on the sensitivity of parameter estimates in the observed process with respect to the latent structure: ignoring time-dependence leads to inefficient estimates of the observed process, highlighting that the linear predictor is sensitive to perturbations in the latent structure; that is, altering the latent structure may produce a significant loss of efficiency of the model parameter estimators. It also shows that, for {\it short} panels, differences between the two approaches are less evident. In terms of efficiency, sensible differences can be observed for {\it large} $T$.

\subsection{A joint model simulation scheme}
In order to assess the empirical performance of the conditional estimator that considers the distribution of ${\bf y}_i$ conditional on the missingness indicator in approximating the joint probability of response and missingness processes, we simulate data from a modified version of the latent class model described in Beunckens et al. (2008). The design of the simulation study is the following. We simulated samples from a bivariate model where the primary response and the dropout variable (namely, $R_{it}$) are conditionally independent given the latent structure. The primary response is Bernoulli distributed with 

\begin{equation}\label{joint1}
g(E[Y_{it}\mid {u}_{it},x_{it}]) = x_{it}'\boldsymbol{\beta}+ {u}_{it}
\end{equation}
where covariate values are again randomly generated from a $N(0,1)$ distribution with fixed parameter $ \beta = 0.5$ and class-specific intercepts equal to $1.0$ and $-1.5$ respectively..
The dropout model can be formulated as follows

\begin{equation}\label{do1}
g(E[R_{it}\mid {u}_{it}]) = \left\{\begin{array}{cc}-3, & j=1\\-1.5 &j=2\end{array}\right.
\end{equation}

with $R_{it}=1, \forall t> t^*: R_{it^*}=1$ and, accordingly, the dropout indicator $S_i$ is defined.
An homogeneous latent Markov structure is considered with $\pi_{ij} = \pi_j = [0.6, 0.4]$ and 

$$
\boldsymbol{\pi} = 
\left[\begin{array}{cc}
\pi_{11} & \pi_{12}\\
\pi_{21} & \pi_{22}
\end{array}\right] = 
\left[\begin{array}{cc}
0.8 & 0.2\\ 0.2 & 0.8
\end{array}\right]
$$

For each of the generated samples, we estimated the parameter values for the two models $M1-M2$ defined before. We give in Table 2 the parameter estimates for the fixed parameter $\beta$ for these models, with standard errors and mean square error, all averaged over simulations.

A clear and consistent path with respect to all conditional models fitted can be observed, indirectly confirming conclusions drawn by Follman and Wu (1995) in a different setting. The modest additional amount of computing resources needed to estimate the conditional model based on a latent Markov heterogeneity structure, and the better results obtained in terms of both bias and mean square error, seem to suggest the use of this new approach, in the binary case discussed here. 

For the classic discrete random effect model, the estimated parameter seems to be persistently biased showing no changes with increasing sample size, while the mean square error seems to slightly increase with increasing number of (maximum) measurements per subjects. The empirical results provide evidence that, in this context, misspecification of the latent distribution may cause bias in the fixed parameter estimates.

\section{Illustrative example}
To illustrate the proposed approach, we consider a dataset derived from a skin cancer longitudinal study conducted to evaluate the effects of $\beta$-carotene to prevent non-melanoma skin cancer in high-risk subjects. After a description of the dataset, we analyze the probability of recording a new cancer with respect to a set of observed covariates. We are going to provide results for the dynamic latent class models proposed in Section 3 in comparison with the proposal described in Alf\`{o} and Aitkin (2000), where the latent process is assumed time-constant. We select such a dataset because the subjects are followed for a limited number of times, making the {\it standard} random effect approach attractive and competitive with respect to the latent Markov-based model introduced in previous sections.

\subsection{Data description}
\begin{figure}
\centering
\centering{\includegraphics[scale=.5]{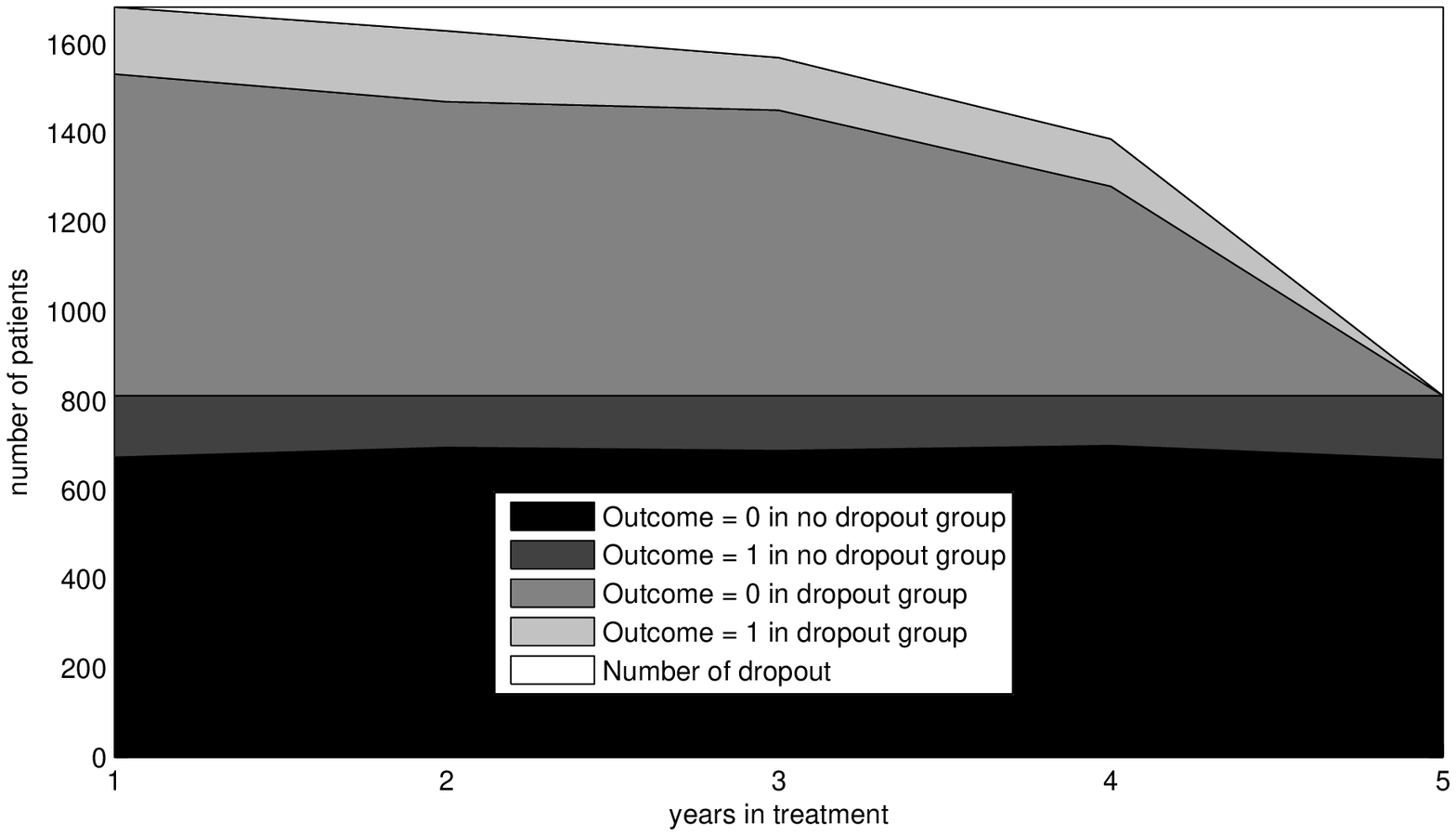}}
\caption{Observed drop-out as a function of time}\label{f:drop}
\end{figure}

Greenberg et al. (1990) presented a randomized, double-blind, placebo-controlled clinical trial of beta-carotene to prevent non-melanoma skin cancer in high risk subjects.
The dataset is a sample of $n=1683$ patients randomized to either placebo or 50mg of beta-carotene per day and observed over a 5-years follow-up, giving a total of 7081 observations. Subjects were examined once a year and biopsied if a cancer was suspected to determine the number of new skin cancers occurring since the last exams.
In addition to the response variable (in our analysis, the presence of a new skin cancer), some covariates are also available: age at the baseline (in years), skin type (1 = burns, 0=otherwise), gender (1 = male, 0 = female), exposure (baseline number of previous skin cancers) and
treatment (1 = $\beta$-carotene, 0 = placebo). Summary statistics over times are provided in Table 3. The number of subjects who dropped out at each time is displayed in Figure \ref{f:drop}. The figure reveals the dropout rate over time. Only 811 subjects complete the study. The 52\% of the subjects dropped out before the completion time. In detail, 53, 60, 183 and 576 subjects were dropped out after first, second, third and fourth year, respectively. Table 4 reports the descriptive statistics among patients in the dropout and non-dropout groups cross-classified with end and non-end presence of a new skin cancer. Those subjects who dropped out have higher percentage of having new skin cancers (17.61\% vs. 15.59\%) as well as end new cancers (56.58\% vs. 46.15\%) than those who completed the study.

\begin{figure}
\centering{\includegraphics[scale=.5]{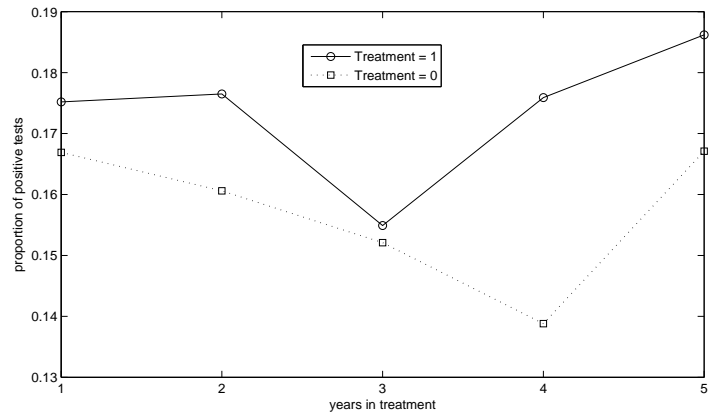}}
\caption{Proportion of new skin cancer by treatment}\label{f:newtreat}
\end{figure}

\begin{figure}
\centering{\includegraphics[scale=.5]{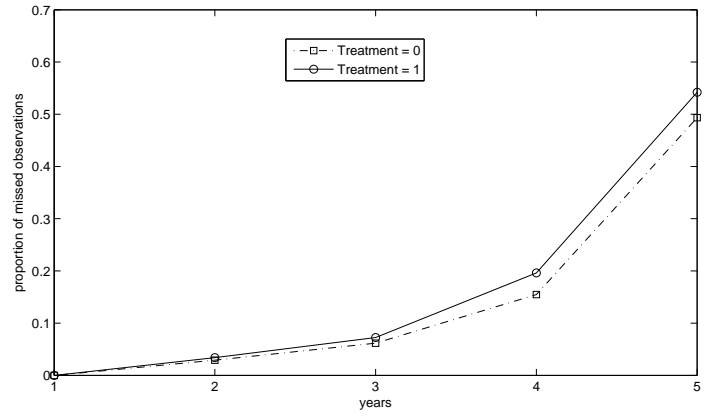}}
\caption{Proportion of missing observations by treatment}\label{f:mistreat}
\end{figure}

The aim of the study was to test the possible cancer-preventing effects of beta carotene. Beta carotene has been considered as a possible treatment associated with a decreasing risk of skin cancer. Thus, the treatment was randomly assigned to subjects who had had a skin cancer by daily administrating either placebo or beta carotene and by yearly looking at the effects of the treatment on the occurrence of new skin cancers. We provide some descriptive info on these two groups of subjects in Figures \ref{f:newtreat}-\ref{f:mistreat}. Treated subjects seem to be more likely to have new skin cancers, while dropouts seems to be similar in the beta carotene and the placebo groups.

\subsection{Model selection and Sensitivity analysis}
Clearly, neither the proposed models nor any other alternative can be seen as a tool to definitely test
for non-ignorable missing mechanism; in fact, without additional information, identification is driven by unverifiable assumptions. Owing to the inclusion of a latent structure in model specification, sound strategies are considered and compared to model dependence on the missing-data. We look at the stability of results across such range of models as an indication
about the belief that we can be put into them.

Results of previous fitting (see e.g. Hasan et al., 2009; Maruotti 2011a) suggested that there is a substantive between-subjects heterogeneity as well as the need of properly accounting for dropouts. We consider several models under different assumptions on the latent structure and on its relation to the time to dropout. In our illustrative example, observations are realizations of Bernoulli random variables with canonical parameter $\lambda_{it}$. Formally, we have 

$$\pi_{ij}(S_i)= \frac{\exp(\gamma_{0j}+\gamma_{1j}s_i)}{1+\sum_{h=1}^{J-1}\exp(\gamma_{0h}+\gamma_{1h}s_i)},$$

$$
\pi_{ikj}(S_i) = \frac{\exp(\phi_{0kj}+\phi_{1kj}S_i)}{1+\sum_{h=1}^{J-1}\exp(\phi_{0kh}+\phi_{1kh}S_i)}
$$

\noindent and the linear predictor is given by

$${\rm logit}(\lambda_{it}) = \beta_0+\beta_1age_{i}+ \beta_2skin_i+\beta_3gender_i+\beta_4exposure_i+\beta_5treatment_i+ u_{it}.$$

Again we aim at comparing our approach with existing methods, and thus we compare our proposal with the one of Alf\`{o} and Aitkin (2000) where 

$$\pi_{ij} = \Pr(u_{i}=u_j\mid S_i) = \frac{\exp(\gamma_{0j}+\gamma_{1j}s_i)}{1+\sum_{h=1}^{J-1}\exp(\gamma_{0h}+\gamma_{1h}s_i)}$$

\noindent and 

$${\rm logit}(\lambda_{i}) = \beta_0+\beta_1age_{i}+ \beta_2skin_i+\beta_3gender_i+\beta_4exposure_i+\beta_5treatment_i+u_{i}.$$

We would remark that, due to imposed restrictions on the latent structure, fitting the random effect model with time-constant heterogeneity could not be accomplished by using the algorithm described in Section 4. A detailed description of algorithms used in this illustrative example to fit it can be found in Aitkin (1996). 

Clearly, our analysis assumes that $J$ is known, whereas in practice it is not. We approach this question in the usual way by selecting different values of $J$ and then select the best solution. We fit the models for increasing values of the number of states (or mixture components) and compare results in terms of penalized likelihood. AIC (Akaike, 1973) and AIC3 (Bozdogan, 1994) are measures of model complexity associated with some criteria that only depend on the number of parameters; some other measures depend on both the number of parameters and the sample size as AICc (Hurvich and Tsai, 1989), AICu (McQuarrie et al., 1997) and BIC (Schwarz, 1978). The dynamic approach proposed in this paper overtakes its respective time-invariant competitor. In this illustrative example, a conditional model based on a latent Markov heterogeneity structure with three classes is preferred according to all the considered AIC-based criteria (see Table 5). BIC, instead, highlights the lack of parsimony of assuming a Markovian-based latent process, making reasonable to consider the more parsimonious time-constant approach. 

Nevertheless, we would remark that different strategies can be pursued to determine the order of the model. Selecting a model that minimizes the penalized likelihood-based indexes provides a parsimonious model that fits the data well. However, in specific application context, it could be preferable to include a larger number of classes at the expense of higher penalized likelihood values. Thus such criteria could be used as a guide only. The final decision on how many classes and which model are to be considered could be evaluated in terms of physical meaning and model interpretability. According to Bartolucci et al. (2009), we suggest the use of the penalized likelihood criteria together with that of diagnostic statistics measuring goodness-of-classification. In fact, as a by-product of the adopted estimation procedures, it is possible to classify subjects on the basis of the posterior probabilities estimates, i.e. $\hat{\xi}_{itj} = {\mathbb E}(u_{itj}\mid {\bf y}_i,\hat{\boldsymbol{\theta}}_r)$ where ${\mathbb E}(u_{itj}\mid {\bf y}_i,\hat{\boldsymbol{\theta}}_r)$ represents how likely the $i$-th subject belongs to class $j$ at time $t$, taking into account the observed process and the time to dropout. This represents a substantial difference with the conclusion that can be drawn adopting a standard parametric approach. Since the obtained classification in latent classes depends on the number of observed measurements, the latent structure could be then used to assess the dependence between the missingness and the outcome processes. To check for missclassification and fuzziness of the obtained classification, we can measure the quality of the classification by the index
$$H = \frac{\sum_{i=1}^n\sum_{t=1}^{S_i}\left(\max\left(\hat{\xi}_{it1},\dots,\hat{\xi}_{itJ}\right)-\frac{1}{J}\right )}{(1-\frac{1}{J})\sum_{i=1}^nS_i}$$

Index $H$ is always between 0 and 1, with 1 corresponding to the
situation of absence of uncertainty in the classification, since one of such posterior
probabilities is equal to 1 for every individual at every time, with all the other probabilities equal
to 0. It helps in identifying if the population clusters are sufficiently well separated. It is worth noting that each state is
characterized by homogeneous values of estimated random effects; thus, conditionally on observed
covariates values, subjects from that state have a similar propensity to the event of interest. Again, our proposal is the preferred model, providing the best goodness-of-classification (see Table 6).

\subsection{Results}
Table 7 presents the results from fitted models under different assumptions on the latent distribution. 
Looking at model parameter estimates and considering a significance level of 5\%, the covariates $age$, $gender$, $skin$ and $exposure$
have a strong influence on the probability of observing a new skin cancer. We also get a
quite unexpected result: the treatment does not affect the observed process, thus posing
questions on program effectiveness. However, such a result is in line with Greenberg et al. (1990).

Fixed effects estimates are similar across models, even if lower parameter values are estimated for $gender$ and $exposure$ if a time-constant latent structure is assumed. Again, this is not surprising and it is in line with the suggestions derived from the simulation study.
In detail, subjects with a burned skin are more likely to have a new skin cancer than other subjects.
The main risk factor for skin cancer is exposure to sunlight, but there are also other factors which may increase the chance of getting a skin cancer. People who have had at least one severe, blistering sunburn are at increased risk of skin cancer. Although people who burn easily are more likely to have had sunburns as a child, sunburns during adulthood also increase the risk of skin cancer.

This is confirmed by looking at fixed effect associated with the age variable. The risk of new skin cancers is higher for the elderly. It is well-known that an age-related reduction of cutaneous melanocytic
density results in more extensive penetration of UV light
into the dermis of elderly subjects, thus causing more
extensive damage. Likewise, age has a negative effect on the
number and function of Langerhans cells of the epidermis,
which are responsible for cutaneous immune function. It is most probable that this immune deficiency is
responsible for the clinical expression of malignancy (see e.g. Swift et al., 2001).

There appears to be a gender
related difference in the development of new skin cancers as
well. Epidemiological studies have
reported the development of significantly more
non-melanoma skin cancer in men than women (Graells, 2004). It is currently believed that lifestyle
choices play a major role in this gender disparity.
Historically men tended to have occupations
that required them to spend more time out in the
sun and overall men are less likely to use sun protection
than women. Occupational exposure to
ultraviolet light does show a strong relationship
between cumulative sunlight exposure and skin cancer risk (Gawkrodger, 2004).
Finally, as expected, exposure (i.e. the number of previous skin cancers) contributes to increment the probability of observing a
new skin cancer. 

The latent structure deserves further
comments. Introducing a latent structure in model specification allows us to define different risks of relapses associated with latent states. In fact, the latent states may be interpreted as the vulnerability
of the subject to recording a new cancer if faced with a high risk situation, and the presence/absence of new cancer on a given year is affected by the subject's vulnerability to
cancer on the given year and whether the subject faced
a high-risk situation on that year. Interpreting the models this way
suggests that relapse is not necessarily an observable condition; it may be a specific latent condition that presumably leads to cancer with higher probability than others, but not with certainty.

\begin{figure}
\centering{\includegraphics[scale=.5]{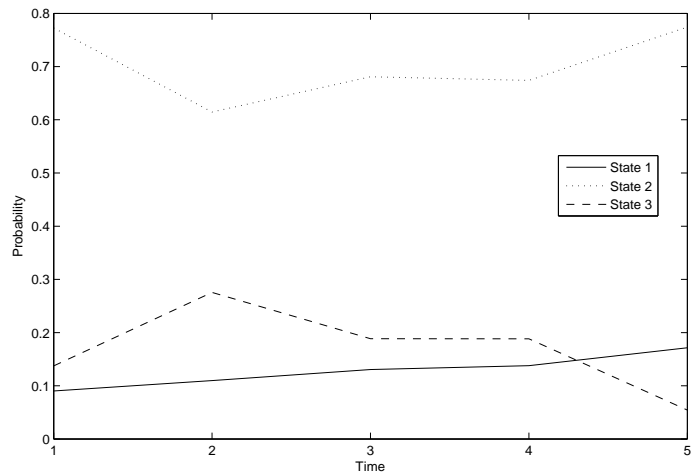}}
\caption{Estimated average probability of each latent state at every
time occasion.}\label{averageprob}
\end{figure}

A three class model is selected according to all AIC-based criteria. Class-specific location estimates are given by \{6.13, -6.01, 4.59\}. Class 2 is clearly the non-cancer class, while both the others imply a more likely vulnerability towards recording a new skin cancer. Latent parameters estimates are provided in Table 8. While the dropout indicator does not affect the probability of belonging to one of the classes at the first time, it is significant in determining the evolution over time and across classes of the propensity of being more likely to record a new skin cancer. Through a local decoding procedure it is possible to draw conclusions on the relationship between vulnerability and dropouts. In fact, we are able to estimate the most likely latent class in correspondence of the last subject observation before attrition. As a result we identify different patterns. Indeed, the 53 subjects dropping out after one visit only are mainly in good health conditions, as well as those 183  whose we recorded three measurement. A different situation has been detected for subjects having two or four measurements. These 60 and 576 subjects, respectively, are mainly clustered in the more vulnerable states (see Table 9). Further, the average probability
of each latent state at every time occasion is represented in
Figure \ref{averageprob}.

\section{Discussion}
\label{discussion}
We introduce a conditional model based on a latent Markov heterogeneity structure for longitudinal binary responses to account for nonignorably dropout mechanism. Through its structure, the model captures unobserved time-dependent heterogeneity between latent subpopulations. We have empirically demonstrated (i.e. in a simulation study) that a right assumption for the latent structure is crucial for drawing valid inferences; indeed, a wrong assumption, as the simulation study shows, can have devastating effects. Therefore a more elaborated approach than the standard random effect approach is required to deal with time-dependence. 

We provide a general framework to deal with heterogeneous
dynamic latent processes. The approach has been defined
using a first order Markov chain, but can be straightforwardly
generalized to higher order Markov chains, the only relevant
change being the increased complexity of the resulting likelihood
and the increased computational effort needed. Moreover, it represents a computationally feasible procedure to
deal with high dimensional random parameters vectors, since the required computational effort is only linear in the number of
dimensions. The proposed approach is an extensions of random effects models and shared parameter models. Our proposal is more general than shared parameter models in that the unobserved heterogeneity not assumed to be independent, but rather to have a time-dependent structure. One consequence of this assumption is that the observed data are also correlated. A drawback of our proposal is that it implies an MNAR mechanism by construction, as shown in a similar context by Tsonaka et al. (2009).

The conditional approach can be further developed re-expressing it to allow for different state-specific intercepts, conditionally on the values assumed by $S_i$. It is particularly convenient to consider such an extension whenever the latent process distribution can be thought of as heavily determined by the conditioning of $S_i$. Of course, our approaches could deal with counts or continuous response data (it requires the use of a different kind of distribution from the Bernoulli used here). 

An important aspect in the specification of the latent structure is the definition of the number of latent states. In this paper, we addressed this issue by looking at several penalized likelihood-based criteria. Nevertheless, further investigation should be devoted to the choice of the number of latent classes. The problem regarding which criterion is able to identify the model with the best trade-off between fit and complexity remains still open.

In our development, all subjects are assumed to have $T$ measurements. In some settings, subjects may have different sequence lengths by design, i.e. informative and noninformative missing arise. A proper summary measure to deal with the missingness mechanism can be obtained by including an indicator variable for the noninformative measurements. A further extension may deal with the distinction between time-dependence and heterogeneity due different dropout propensities. In other words, ideally, unobserved heterogeneity due to dropout could be disentangled from {\it behavior}-persistence. 

As a final remark, we would point out that the proposed approach can be extended to the continuous-time case, i.e. when the time a subject spends in the study is continuous. Then the observed and the dropout processes can be jointly modeled in a unique framework as recently proposed, in a survival context, by Viviani et al. (2013).

\section*{Appendix}
Let $$I_t = \{i:S_i=t; i = 1,\dots,n\}$$ be the subset of subjects having $S_i=t$ with $$|I_1|+|I_2|+\dots+|I_T|=n_1+n_2+\dots+n_T=n.$$

Now the log-likelihood can be explicitly written as follows

\begin{eqnarray*}\label{like1bis}
\log L(\cdot\mid S_i) &=&
\sum_{i\in I_1}\log\left\{\sum_{{\bf u}_{i1}}\cdots \sum_{{\bf u}_{iT}}\left[\Pr({\bf u}_{i1}\mid S_i)\prod_{t=2}^T\Pr({\bf u}_{it}\mid {\bf u}_{it-1}, S_i)\prod_{t=1}^Tf(y_{it}\mid {\bf u}_{it})\right]\right\}\\&+&\dots\\&+&\sum_{i\in I_T}\log\left\{\sum_{{\bf u}_{i1}}\cdots \sum_{{\bf u}_{iT}}\left[\Pr({\bf u}_{i1}\mid S_i)\prod_{t=2}^T\Pr({\bf u}_{it}\mid {\bf u}_{it-1}, S_i)\prod_{t=1}^Tf(y_{it}\mid {\bf u}_{it})\right]\right\}
\end{eqnarray*}

Differentiating the previous equation with respect to hidden chain parameters under the constraints $\sum_j \pi_{ij}(S_i=t) = 1$ and 
$\sum_j\pi_{ikj}(S_i=t)=1$ and equating to zero the corresponding derivatives, the M-step reduces to

$$\hat{\pi}_{ij}(S_i=t) = \sum_{i\in I_t}\frac{\hat{\xi}_{itj}(S_i=t)}{n_t}$$
and 
$$\hat{\pi}_{ikj}(S_i=t) = \sum_{i\in I_t}\frac{\hat{\zeta}_{itkj}(S_i=t)}{\sum_j\hat{\zeta}_{itkj}(S_i=t)}$$
where $\hat{\xi}_{itj}(S_i=t)$ represents the posterior probability that subject $i$, with $S_i=t$, comes from state $j$ (similarly for $\hat{\zeta}_{itkj}(S_i=t)$
).

\section*{Acknowledgment}
I gratefully acknowledge Dr Robert Greenberg for kindly providing the data set on the skin cancer prevention study. I thank Dankmar B\"ohning for his comments on a draft version of this paper.

\newpage


\begin{table} \label{t:sim1}
\caption{Simualtion results: Conditional simulation scheme. Bias, Standard deviation and mean square error of the fixed effect $\beta$}
\centering
\begin{tabular}{cccc||ccc}
  & \multicolumn{3}{c||}{$n = 100\quad T = 5$}&  \multicolumn{3}{c}{$n = 100\quad T = 10$}\\
 
& Bias & Std. Dev. & MSE &  Bias & Std. Dev. & MSE \\
 
M1 & 0.1125 & 0.1840 & 0.0465 &  0.0250 & 0.1100 & 0.0127\\
M2 & -0.1288 & 0.0762 & 0.0224  & -0.1085 & 0.0695 & 0.0166\\
 
 \hline\hline
  & \multicolumn{3}{c||}{$n = 250\quad T = 5$}&  \multicolumn{3}{c}{$n = 250\quad T = 10$}\\
 
& Bias & Std. Dev. & MSE &  Bias & Std. Dev. & MSE \\
 
M1 & 0.0729 & 0.1219 & 0.0202 & 0.0082 & 0.0670 & 0.0046\\
M2 & -0.1256 & 0.0487 & 0.0182 & -0.1073 & 0.0400 & 0.0131\\
 \hline\hline
 
  & \multicolumn{3}{c||}{$n = 500\quad T = 5$}&  \multicolumn{3}{c}{$n = 500\quad T = 10$}\\
 
& Bias & Std. Dev. & MSE &  Bias & Std. Dev. & MSE \\
 
M1 &  0.0546 & 0.0647 & 0.0072 & -0.0071 & 0.0405 & 0.0017\\
M2 & -0.1202 & 0.0336 & 0.0156 & -0.1097 & 0.0281 & 0.0128\\
 
\end{tabular}
\end{table}

\begin{table}
\caption{Simualtion results: Joint model simulation scheme. Bias, Standard deviation and mean square error of the fixed effect $\beta$}
\centering
\begin{tabular}{cccc||ccc}
 \label{t:joi1}
  & \multicolumn{3}{c||}{$n = 100\quad T = 5$}&  \multicolumn{3}{c}{$n = 100\quad T = 10$}\\
 
& Bias & Std. Dev. & MSE &  Bias & Std. Dev. & MSE \\
 
M1 & 0.076 & 0.143 & 0.026 & 0.048 & 0.124 & 0.018\\
M2 & -0.065 & 0.084 & 0.011 & -0.108 & 0.061 & 0.015\\
 \hline\hline
 
  & \multicolumn{3}{c||}{$n = 250\quad T = 5$}&  \multicolumn{3}{c}{$n = 250\quad T = 10$}\\
 
& Bias & Std. Dev. & MSE &  Bias & Std. Dev. & MSE \\
 
M1 &  0.047 & 0.096 & 0.011 & -0.003 & 0.077 & 0.006\\
M2 & -0.088 & 0.042 & 0.010 & -0.118 & 0.038 & 0.016\\
 
 \hline\hline
  & \multicolumn{3}{c||}{$n = 500\quad T = 5$}&  \multicolumn{3}{c}{$n = 500\quad T = 10$}\\
 
& Bias & Std. Dev. & MSE &  Bias & Std. Dev. & MSE \\
 
M1 & 0.007 & 0.055 & 0.003 & 0.008 & 0.051 & 0.003\\
M2 & -0.097& 0.032& 0.010 & -0.107 & 0.030 & 0.012\\
 
\end{tabular}
\end{table}

\begin{landscape}\begin{table}\caption{Skin cancer data. Summary statistics}\footnotesize
\begin{center}
\begin{tabular}{c|cc|cc|cc|cc|cc}\label{t:des}
Variables & \multicolumn{2}{c|}{First} & \multicolumn{2}{c|}{Second}& \multicolumn{2}{c|}{Third}& \multicolumn{2}{c|}{Fourth}& \multicolumn{2}{c}{Fifth}\\
 & \multicolumn{2}{c|}{Occasion} & \multicolumn{2}{c|}{Occasion}& \multicolumn{2}{c|}{Occasion}& \multicolumn{2}{c|}{Occasion}& \multicolumn{2}{c}{Occasion}\\
\hline
 & Mean & Std.Dev  & Mean & Std.Dev  & Mean & Std.Dev  & Mean & Std.Dev  & Mean & Std.Dev \\
\hline
\% new cancer & 0.171 & 0.142 & 0.169 & 0.140 & 0.154 & 0.130 & 0.157 & 0.132 & 0.176 & 0.145\\
Age & 63.022 & 9.940 & 63.005 & 9.927 & 62.890 & 9.941 & 63.089 & 9.784 & 62.975 & 9.731\\
Gender & 0.689 & 0.463 & 0.684 & 0.465 & 0.678 & 0.467 & 0.680 & 0.467 & 0.660 & 0.474\\
Skin & 0.456 & 0.498 & 0.454 & 0.498 & 0.449 & 0.498 & 0.450 & 0.498 & 0.433 & 0.496\\
Exposure & 2.892 & 3.409 & 2.868 & 3.379 & 2.838 & 3.318 & 2.795 & 3.305 & 2.835 & 3.395\\
Treatment & 0.509 & 0.500 & 0.507 & 0.500 &0.506 & 0.500 & 0.496 & 0.500 & 0.483 & 0.500\\
\hline
\# subjects & \multicolumn{2}{c|}{1683}& \multicolumn{2}{c|}{1630}& \multicolumn{2}{c|}{1570}& \multicolumn{2}{c|}{1387}& \multicolumn{2}{c}{811}
\end{tabular}
\end{center}
\end{table}\end{landscape}

\begin{table}\caption{Skin cancer data. Cross-tabulation: dropout/non-dropout groups vs. end-use}
\begin{center}
\begin{tabular}{c|c|cc|c}\label{t:crosstab}
&Variables& cancer at  &cancer at  & Total\\
&& the last time = 1 &the last time = 0 & \\
\hline
Dropout ($T_i<5$) & & &  \\
\hline
& Skin Cancer = 1 & 0.57 & 0.09 & 0.18\\
& Average Age & 63.83 & 62.84 & 63.02\\
& Average Exposure & 5.40 & 2.29 & 2.87\\
&  Skin=1 & 0.59 & 0.45 & 0.47\\
&  Treatment=1 & 0.59 & 0.51 & 0.53\\
&  Gender=1 &  0.85&0.68&0.71\\
\hline
Non Dropout ($T_i=5$) & & &  \\
\hline
& Skin Cancer = 1 & 0.46 & 0.09 & 0.16\\
& Average Age & 64.61 & 62.62 & 62.97\\
& Average Exposure & 4.72 & 2.43 & 2.83\\
&  Skin=1 & 0.50 & 0.42 & 0.43\\
&  Treatment=1 & 0.51 & 0.48 & 0.48\\
&  Gender=1  & 0.78 & 0.63 & 0.66\\
\hline
Total & & &  \\
\hline
& Skin Cancer = 1 & 0.51 & 0.09 & 0.16\\
& Average Age & 64.27 & 62.71 & 63.00\\
& Average Exposure & 5.01 & 2.37 & 2.85\\
&  Skin=1 & 0.54 & 0.43 & 0.45\\
&  Treatment=1&0.55 & 0.49 & 0.50\\
&  Gender=1 & 0.81 & 0.65 & 0.68
\end{tabular}
\end{center}\end{table}

\begin{landscape}\begin{table}\caption{Model selection}
\label{t:AIC}\small
\centering	
\begin{tabular}{c|cccccccc}

Model&	AIC	&	AIC3	&	AICc	&	AICu	&	BIC &	Log-	&	\# of  & 	\# of states \\
&	&	&	&	&&likelihood	& param. & 	\\\hline

\hline\hline
Alf\`{o} and Aitkin (2000)\\
Time-constant random effects \\
&	5570.320	&	5579.320	&	5570.428	&	5580.457	&	5619.175&-2776.160	&	9	&2\\
&	5564.694	&	5576.694	&	5564.881	&	5577.931	&	5629.834&-2770.347	&	12	&3\\
&	5570.434	&	5585.434	&	5570.722	&	5586.798	&	5651.859&-2770.217	&	15	&4\\
\hline \hline
Our Proposal\\
Time-varying random effects\\
&5555.799 & 5568.799 & 5556.017 & 5570.076 & 5626.368 & -2764.899	&	13	& 2\\
&5499.998 & 5523.998 & 5500.722 &5525.909&5630.278&-2726.000&	24	&3\\
&5505.499 & 5544.499& 5507.398&5547.881&5718.204&-2713.750&	39	&4\\

\end{tabular}
\end{table}
\end{landscape}

\begin{table}
\caption{Goodness-of-classification}
\label{t:H}
\begin{center}
\begin{tabular}{c|c}
Model&	$H$\\
\hline\hline
Time-constant random effects & 0.3989\\
($J = 3$)\\
\hline \hline
Time-varying random effects & 0.4467\\
($J = 3$)\\
\end{tabular}
\end{center}
\end{table}

\begin{table}
\caption{Parameters estimates: observed process}
\label{t:res}\small
\begin{center}
\begin{tabular}{c||cc||cc}
Variables & \multicolumn{2}{c||}{Time-constant random effects}  & \multicolumn{2}{c}{Time-varying random effects}\\
 & \multicolumn{2}{c||}{Alf\`{o} and Aitkin (2000)}  & \multicolumn{2}{c}{Our proposal}\\\hline
\hline
& Coeff. & Std. Dev& Coeff. & Std. Dev\\\hline
Age & 0.022 & 0.005& 0.021 & 0.005 \\
Skin & 0.308 & 0.098 & 0.398 & 0.088\\
Gender &  0.774 & 0.112 & 0.935 & 0.101\\
Exposure &  0.228 & 0.017 & 0.569 & 0.026\\
Treatment & 0.084 & 0.096 & 0.090 & 0.087\\
Intercept & -4.938 & 0.434 & -9.738 & 0.497\\
\hline
\end{tabular}
\end{center}
\end{table}
\begin{table}
\caption{Time-varying random effects model: hidden parameter estimates}
\label{hidden}
\begin{center}
\begin{tabular}{c||cc}
Parameter& Coeff. & Std. Dev.\\
\hline
$\gamma_{01}$&-0.509 & 0.310 \\
$\gamma_{11}$&-0.100 & 0.071\\
$\gamma_{02}$& 0.276&0.236\\
$\gamma_{12}$&-0.029&0.056\\
\hline\hline
$\phi_{011}$ & -10.558 & 1.009\\
$\phi_{111}$ & 4.507 & 0.367\\
$\phi_{012}$ &-15.259 & 1.076\\
$\phi_{112}$ & 5.094 & 0.376\\
$\phi_{021}$ & -7.856 & 0.603\\
$\phi_{121}$ & 1.317 & 0.126\\
$\phi_{022}$ & -2.664 & 0.199\\
$\phi_{122}$ & 0.665 & 0.044\\
$\phi_{031}$ & -10.065 & 2.045\\
$\phi_{131}$ & 1.267 & 0.423\\
$\phi_{023}$ & -1.080 & 0.288\\
$\phi_{123}$ & 0.341 & 0.041
\end{tabular}
\end{center}
\end{table}

\begin{table}
\caption{Time-varying random effects model. Local decoding: Clustering at attrition}
\label{exit}\small
\begin{center}
\begin{tabular}{c||ccccc}
& \multicolumn{5}{c}{$S_i$}\\
State & 1 & 2 & 3 & 4 & 5\\
\hline
1 & 5 & 0 & 25 & 55 & 139\\
2 & 39 & 6 & 143 & 297 & 628\\
3 & 9 & 54 & 15 & 224 & 44
\end{tabular}
\end{center}
\end{table}

\end{document}